\documentclass[conference]{IEEEtran}
\IEEEoverridecommandlockouts

\usepackage{cite}
\usepackage{amsmath,amssymb,amsfonts}
\usepackage{algorithmic}
\usepackage{graphicx}
\usepackage{textcomp}
\usepackage{listings}
\usepackage[bookmarks=false]{hyperref}
\usepackage{makecell}
\usepackage{enumitem}
\usepackage{mathtools}
\usepackage{supertabular,booktabs}
\usepackage[super]{nth}
\usepackage{tikz}
\usetikzlibrary{shapes,positioning,calc}
\def\BibTeX{{\rm B\kern-.05em{\sc i\kern-.025em b}\kern-.08em
    T\kern-.1667em\lower.7ex\hbox{E}\kern-.125emX}}
\begin{document}

\newlist{legal}{enumerate}{10}
\setlist[legal]{label*=\arabic*.}

\lstset{
    language=C++,
    breaklines = true,
    upquote = true,
    columns = flexible,
    basicstyle = \ttfamily,
    frame = single,
    keepspaces = true,
}
\newcommand{\defn}{\stackrel{\textrm{\clap{\tiny def}}}{=}}

\title{Securing the EDK~II Image Loader
}

\author{\IEEEauthorblockN{Marvin Häuser}
\IEEEauthorblockA{\textit{Technische Universität Kaiserslautern;}\\\textit{Ivannikov Institute for System Programming}\\\textit{of the Russian Academy of Sciences} \\
Kaiserslautern, Germany \\
mhaeuser@posteo.de}
\and
\IEEEauthorblockN{Vitaly Cheptsov}
\IEEEauthorblockA{\textit{Ivannikov Institute for System Programming}\\\textit{of the Russian Academy of Sciences}\\
Moscow, Russia \\
cheptsov@ispras.ru}
}

\maketitle

\begin{tikzpicture}[remember picture, overlay]
    \path node at ($(current page.south) + (0,0.65in)$) {
        \begin{minipage}{\textwidth} \footnotesize
            M. Häuser and V. Cheptsov, "Securing the EDK II Image Loader," 2020 Ivannikov Ispras Open Conference (ISPRAS), 2020, pp. 16-25, DOI: \href{https://doi.org/10.1109/ISPRAS51486.2020.00010}{10.1109/ISPRAS51486.2020.00010}.

            \copyright~2020 IEEE. Personal use of this material is permitted. Permission
            from IEEE must be obtained for all other uses, in any current or future media,
            including reprinting/republishing this material for advertising or promotional
            purposes, creating new collective works, for resale or redistribution to
            servers or lists, or reuse of any copyrighted component of this work in other
            works.
        \end{minipage}
    };
\end{tikzpicture}
\begin{abstract}

The Unified Extensible Firmware Interface~(UEFI) is a standardised interface between the firmware and
the operating system used in all x86-based platforms over the past ten years, which continues to spread
to other architectures such as ARM and RISC-V. The UEFI incorporates a modular design based on
images containing a driver or an application in a Common Object File Format~(COFF) either
as a Portable Executable~(PE) or as a Terse Executable~(TE). The de-facto standard generic UEFI services
implementation, including the image loading functionality, is TianoCore EDK~II. Its track of
security issues shows numerous design and implementation flaws some of which are yet to be addressed. In this paper
we outline both the requirements for a secure UEFI Image Loader and the issues of the existing implementation.
As an alternative we propose a formally verified Image Loader supporting both PE and TE images with
fine-grained hardening enabling a seamless integration with EDK~II and subsequently with the other firmwares.

\end{abstract}


\section{Introduction}

The fast spread of the Unified Extensible Firmware Interface~(UEFI)~\cite{uefispec} over x86-based platforms would not have
been possible without the adoption of other specifications backed with existing widely
available tools. The use of the Portable Executable~(PE)~\cite{pespec} format let Intel reuse the existing specification
for the format itself as well as the established compiler infrastructure. This decision led the UEFI Forum
to keep including the PE format as part of the UEFI specification essentially making it the core of the UEFI
modular architecture.

The UEFI architecture is designed in a way that not all the modules used during the boot sequence
can be equally trusted. While firmware module trust is assumed as a fact by the UEFI and UEFI PI~\cite{pispec}
specifications and is handled by vendor-specific technologies such as Intel BootGuard~\cite{bootguard},
other module protection is not taken for granted and is part of the ``Secure Boot and Driver Signing''
and ``Secure Technologies'' sections of the UEFI specification.

A driver coming with an option ROM, a UEFI application, or a standalone UEFI driver are by all means
less trusted but usually do not execute with significantly reduced privileges compared to firmware drivers and
applications. To circumvent this disadvantage each module external to the firmware is
verified to belong to an authority that is trusted by the firmware prior to execution. The standardised
verification algorithm in UEFI for image signature is Microsoft Authenticode~\cite{pehashspec},
which bundles image signature within the PE file and requires most of structural parsing for its
verification. While the choice of this algorithm made it possible to reuse the code signing tools
also used for Windows driver signing easing the development process, it also imposed a number
of extra security requirements on the software in charge of the signature verification process
on the firmware side.

Today most of the platform-independent code of the UEFI firmwares is implemented as part of
the open-source TianoCore EDK~II~\cite{edk2} toolkit. This implementation consisting of about a million
lines of C code also includes the code to handle the PE images that can be found in practically
any modern firmware. In an attempt to make the implementation more reliable contributions
to EDK~II go through a thorough review process backed with additional specifications
covering the development from architecture to code style, automated testing
on several platforms and on-demand testing with several static and dynamic analysers~\cite{hbfa}.
The amount of vulnerabilities found in modern firmwares over the last ten years
shows that the measures taken are clearly not enough~\cite{fwrootkits}. While the attempts to introduce
other programming languages with built-in safety checkers may improve the situation
with naive security issues like buffer overflows, logic errors are much harder to find
and in general cannot be identified automatically.

In this paper we suggest using formal verification to secure the PE image loading
functionality in UEFI. Section~\ref{subsec:stateofart} describes the issues of the current implementation
present in EDK~II and specifications separating design defects, functional defects, and potential
issues not covered by the existing implementation. Section~\ref{sec:astraver} provides a basic introduction
into formal verification and introduces AstraVer, the tool we used to verify the code against our models.
Section~\ref{sec:implementation} describes the implementation and the verification model.

\emph{\textbf{Note:}} We would like to clarify on our terminology to avoid ambiguities. Please note that specific terms from the cited specifications will not be repeated.
\begin{itemize}
    \item ``[Raw] file'' shall mean the format in which a PE or TE image is saved on storage devices.
    \item ``[Loaded] image'' shall mean the format of a PE or TE image expanded into the execution environment.
    \item ``Verification'' shall mean the process of ensuring the plausibility of data structures within a file or image conforming to the PE format~\cite{pespec}.
    \item ``Hashing'' shall mean the cryptographic hashing of the raw file data in a way described in the used signature verification algorithm such as Authenticode~\cite{pehashspec}.
    \item ``Loading'' shall mean the transformation of a TE or PE raw file into a loaded image.
    \item ``Relocating'' shall mean rebasing a loaded image from one virtual address base to another.
    \item ``Image Loader'' shall mean the library that performs actions such as verification, hashing, loading or relocating of a raw file or image.
\end{itemize}

\section{State of the Art} \label{subsec:stateofart}
The EDK~II Image Loader contains various defects. While some are bugs of the specific
implementation~(``functional defects''), others are fundamental design flaws of the library design that
significantly raise the risk of introducing functional defects~(``design defects'').

\subsection{Design defects}
\begin{enumerate}
    \item \textbf{Time-of-check/Time-of-use vulnerable design}
    \label{desvul:TOCTOU}

    The EDK~II Image Loader uses a function provided by the caller to abstract the gradual image data reading\footnote{
        EDK~II~\cite{edk2} \href{https://github.com/tianocore/edk2/blob/6c8dd15c4ae42501438a525ec41299f365f223cb/MdePkg/Include/Library/PeCoffLib.h\#L33-L70}
        {PeCoffLib.h:33-70}
    }.
    As made clear by the function specification its explicit purpose is to allow arbitrary, untrusted data
    sources such as networks. At the time of writing there are a total of 18 calls to this function across
    multiple stages of loading which makes ensuring data is not re-read and used unverified
    (\href{https://cwe.mitre.org/data/definitions/367.html}{\textit{TOC/TOU}}) a non-trivial task. This
    is currently an actual threat as the image headers are read early for verification\footnote{
        EDK~II~\cite{edk2} \href{https://github.com/tianocore/edk2/blob/6c8dd15c4ae42501438a525ec41299f365f223cb/MdePkg/Library/BasePeCoffLib/BasePeCoff.c\#L81-L86}
        {BasePeCoffLib/BasePeCoff.c:81-86}

        EDK~II~\cite{edk2} \href{https://github.com/tianocore/edk2/blob/6c8dd15c4ae42501438a525ec41299f365f223cb/MdePkg/Library/BasePeCoffLib/BasePeCoff.c\#L112-L117}
        {BasePeCoffLib/BasePeCoff.c:112-117}
    }
    and then again when loading the image into its final memory destination\footnote{
        EDK~II~\cite{edk2} \href{https://github.com/tianocore/edk2/blob/6c8dd15c4ae42501438a525ec41299f365f223cb/MdePkg/Library/BasePeCoffLib/BasePeCoff.c\#L1266-L1271}
        {BasePeCoffLib/BasePeCoff.c:1266-1271}
    }
    without ensuring equivalence or re-verifying the data read. Whether this is an actual security
    vulnerability depends on whether the caller-provided read function uses a trusted source~(e.g. main
    memory) or not~(e.g. network).\\
    This defect is addressed by hardened design~(\ref{sect:hardeneddesign}).

    \item \textbf{Out-of-bounds vulnerable design}
    \label{desvul:OOB}

    HII data that is provided to the caller\footnote{
        EDK~II~\cite{edk2} \href{https://github.com/tianocore/edk2/blob/6c8dd15c4ae42501438a525ec41299f365f223cb/MdeModulePkg/Core/Dxe/Image/Image.c\#L1406-L1416}
        {Core/Dxe/Image/Image.c:1406-1416}
    } is variably-sized. Since its internal offsets are not validated to be in bounds of the image address space
     and it is also not stored with the maximum bounds, so that it can be safely accessed\footnote{
        EDK~II~\cite{edk2} \href{https://github.com/tianocore/edk2/blob/6c8dd15c4ae42501438a525ec41299f365f223cb/MdePkg/Library/BasePeCoffLib/BasePeCoff.c\#L1613}
        {BasePeCoffLib/BasePeCoff.c:1613}
    }, the caller is left at a high risk of an out-of-bounds access.

    This defect is addressed by hardened design~(\ref{sect:hardeneddesign}) and raw file and loaded image models~(\ref{sect:imagemodels}).

    \item \textbf{Untrusted data in a trusted environment}

    This defect closely relates to the previous defect~\ref{desvul:OOB}, but is caller- rather than callee-centric. The UEFI specification requires the UEFI image's HII resource data, which is variable-length\footnote{
        EDK~II~\cite{edk2} \href{https://github.com/tianocore/edk2/blob/6c8dd15c4ae42501438a525ec41299f365f223cb/MdePkg/Include/Uefi/UefiInternalFormRepresentation.h\#L48-L51}
        {UefiInternalFormRepresentation.h:48-51}
    }, to be exposed to the trusted environment~(UEFI protocol installation). Consequently, the Image
    Loader locates and exposes the location of this data\footnote{
        EDK~II~\cite{edk2} \href{https://github.com/tianocore/edk2/blob/6c8dd15c4ae42501438a525ec41299f365f223cb/MdePkg/Library/BasePeCoffLib/BasePeCoff.c\#L1613}
        {BasePeCoffLib/BasePeCoff.c:1613}
    }. However, no verification occurs within or outside the Image Loader before the data is exposed at
    load time\footnote{
        EDK~II~\cite{edk2} \href{https://github.com/tianocore/edk2/blob/6c8dd15c4ae42501438a525ec41299f365f223cb/MdeModulePkg/Core/Dxe/Image/Image.c\#L1406-L1416}
        {Core/Dxe/Image/Image.c:1406-1416}
    }. Loading an image cannot be seen as an act of trusting it, as with a safe loader implementation
    the image did not yet gain or was given any control over the system.\\
    This defect needs to be addressed on UEFI specification level.

    \item \textbf{Ambiguous context ownership}
    \label{desvul:ambiguousctx}

    Traditionally, context instances are owned by the library that declares them --- coming from an
    object-orientated programming paradigm, they can be thought of as non-static members of a class
    instance. While public reading can be tolerated to simplify the interface, writes should be performed
    strictly within the library implementation to preserve consistent states. EDK~II currently relies on
    external context writes as part of the library function contracts\footnote{
        EDK~II~\cite{edk2} \href{https://github.com/tianocore/edk2/blob/6c8dd15c4ae42501438a525ec41299f365f223cb/MdePkg/Include/Library/PeCoffLib.h\#L208-L209}
        {PeCoffLib.h:208-209}
    }.\\
    This defect is addressed by hardened design~(\ref{sect:hardeneddesign}).

    \item \textbf{Complicated function contracts}
    \label{desvul:funcontr}

    Function contracts should be easy to understand and control flows easy to comprehend. Presently EDK~II
    relies on explicitly declaring the context fields required to be valid and the context fields ensured to
    be valid by each function\footnote{
        EDK~II~\cite{edk2} \href{https://github.com/tianocore/edk2/blob/6c8dd15c4ae42501438a525ec41299f365f223cb/MdePkg/Include/Library/PeCoffLib.h\#L199-L201}
        {PeCoffLib.h:199-201}, EDK~II~\cite{edk2} \href{https://github.com/tianocore/edk2/blob/6c8dd15c4ae42501438a525ec41299f365f223cb/MdePkg/Include/Library/PeCoffLib.h\#L208-L209}
        {PeCoffLib.h:208-209}
    }. Except for fields explicitly allowed to be publicly read this severely overcomplicates the
    documentation.\\
    This defect is addressed by hardened design~(\ref{sect:hardeneddesign}).

    \item \textbf{Decentralisation of code}
    \label{desvul:decentralised}

    There are several locations throughout the EDK~II codebase that access PE and TE image structures directly
    instead of utilising the Image Loader\footnote{
        EDK~II~\cite{edk2} \href{https://github.com/tianocore/edk2/blob/6c8dd15c4ae42501438a525ec41299f365f223cb/MdeModulePkg/Core/Dxe/Mem/MemoryProfileRecord.c\#L251-L353}
        {Core/Dxe/Mem/MemoryProfileRecord.c:251-353}
    }. This is strongly undesired especially for the hashing algorithm which is present four times\footnote{
        EDK~II~\cite{edk2} \href{https://github.com/tianocore/edk2/blob/6c8dd15c4ae42501438a525ec41299f365f223cb/SecurityPkg/Library/DxeImageVerificationLib/DxeImageVerificationLib.c\#L273-L293}
        {DxeImageVerificationLib/DxeImageVerificationLib.c:273-293}

        EDK~II~\cite{edk2} \href{https://github.com/tianocore/edk2/blob/6c8dd15c4ae42501438a525ec41299f365f223cb/SecurityPkg/VariableAuthenticated/SecureBootConfigDxe/SecureBootConfigImpl.c\#L1809-L1825}
        {SecureBootConfigDxe/SecureBootConfigImpl.c:1809-1825}

        EDK~II~\cite{edk2} \href{https://github.com/tianocore/edk2/blob/6c8dd15c4ae42501438a525ec41299f365f223cb/SecurityPkg/Library/DxeTpmMeasureBootLib/DxeTpmMeasureBootLib.c\#L267-L300}
        {DxeTpmMeasureBootLib/DxeTpmMeasureBootLib.c:267-300}

        EDK~II~\cite{edk2} \href{https://github.com/tianocore/edk2/blob/6c8dd15c4ae42501438a525ec41299f365f223cb/SecurityPkg/Tcg/Tcg2Dxe/MeasureBootPeCoff.c\#L76-L101}
        {Tcg2Dxe/MeasureBootPeCoff.c:76-101}
    } outside the Image Loader, where the function logically belongs. Without the library explicitly
     expressing the guarantees it makes regarding image data structures~(as opposed to only its own
    context structure) the code cannot trust the structures. Validating them then may lead to code
    duplication depending on the library implementation details. Furthermore, the Image Loader is exposed via two separate libraries\footnote{
        EDK~II~\cite{edk2} \href{https://github.com/tianocore/edk2/blob/6c8dd15c4ae42501438a525ec41299f365f223cb/MdePkg/Include/Library/PeCoffGetEntryPointLib.h}
        {PeCoffGetEntryPointLib.h}
    } for no obvious reason. This decentralisation adds cost to review,
    validation and bug fixing.\\
    This defect is addressed by hardened design~(\ref{sect:hardeneddesign}).
    \pagebreak
    \item \textbf{Centralisation of hashing ciphers}
    \label{desvul:ciphers}

    The hashing function mentioned in ``Decentralisation of code'' statically defines a list of supported
    hashing ciphers\footnote{
        EDK~II~\cite{edk2} \href{https://github.com/tianocore/edk2/blob/6c8dd15c4ae42501438a525ec41299f365f223cb/SecurityPkg/Library/DxeImageVerificationLib/DxeImageVerificationLib.c\#L322-L342}
        {DxeImageVerificationLib/DxeImageVerificationLib.c:322-342}
    }. This means that not only custom cryptography is not easily supported, there also is no easy platform control over
    the allowed ciphers.\\
    This defect is addressed by hardened design~(\ref{sect:hardeneddesign}).

    \item \textbf{Runtime relocation is status-less}
    \label{desvul:statusless}

    Unlike all other APIs, the EDK~II Image Loader does not report a status for the Runtime relocation
    operation\footnote{
        EDK~II~\cite{edk2} \href{https://github.com/tianocore/edk2/blob/6c8dd15c4ae42501438a525ec41299f365f223cb/MdePkg/Include/Library/PeCoffLib.h\#L356}
        {PeCoffLib.h:356}
    }. However, there are obvious and known error conditions where the operation may not succeed and they
    are handled inappropriately by silently returning or ASSERTing\footnote{
        EDK~II~\cite{edk2} \href{https://github.com/tianocore/edk2/blob/6c8dd15c4ae42501438a525ec41299f365f223cb/MdePkg/Library/BasePeCoffLib/BasePeCoff.c\#L1748-L1760}
        {BasePeCoffLib/BasePeCoff.c:1748-1760}
    }.\\
    This defect is addressed by hardened design~(\ref{sect:hardeneddesign}).

    \item \textbf{Data structures are not protected from relocation}
    \label{desvul:reloctrashing}

    Base Relocation targets may point to parts of the image data structures, the most notable of which is the Relocation
    Directory itself. This may corrupt the relocation data making it unusable for further processes such as Runtime relocation.\\
    This defect is addressed by the loaded image model~(\ref{sect:imagemodels}).

    \item \textbf{Unclear relocation semantics}
    \label{desvul:unclearreloc}
    
    The descriptions for the \textit{IMAGE\_REL\_BASED} \textit{LOW} and \textit{HIGH} Base Relocation types  clearly outline they are compound. However, their semantics does not define how the carry for the low component is to be handled. This already has caused confusion in the past leading to a patch that reasonably adds the carry to the high component but that also makes the assumption that those Base Relocation types are always consecutive and ordered\footnote{
        \href{https://www.mail-archive.com/edk2-devel@lists.sourceforge.net/msg16005.html}{www.mail-archive.com/edk2-devel@lists.sourceforge.net/msg16005.html}
    }, which is not guaranteed by the PE format.\\
    This defect is addressed by hardened design~(\ref{sect:hardeneddesign}). It additionally needs to be addressed on PE format~\cite{pespec} level.

    \item \textbf{Image information is leaked}
    \label{desvul:infoleak}

    The image headers, debug information, and Relocation Directory contain information about exact locations such as function addresses in the
    image data. This information can be used to more easily locate gadgets and potentially
    transparently spoof certain operations when write access is gained to the image data.\\
    This defect is addressed by hardened design~(\ref{sect:hardeneddesign}).

    \item \textbf{Nondeterminism}
    \label{desvul:nondet}
    
    It is not required for Base Relocations to be processed in a specific order or to target distinct memory by the PE format. Furthermore, there are no constraints to the first image section (such as that it must be the start of the image), whether the image headers must be loaded or what needs to happen with the possible gap from their end to the beginning of the first section's memory. Hence, valid loaders may produce different results for valid images (nondeterminism).\\
    This defect is addressed by the loaded image model~(\ref{sect:imagemodels}) and ACSL models\footnote{
        They are exclusive to the code and are not discussed in this document.
    }. It additionally needs to be addressed on PE format~\cite{pespec} level.
\end{enumerate}

\subsection{Functional defects}
\begin{enumerate}    
    \item \textbf{Out-of-bounds accesses}
    
    The current \textit{PeCoffLoaderImageAddress} function returns a pointer to a requested offset within the
    image buffer. However, it does not return the remaining number of bytes to the end of the buffer\footnote{
        EDK~II~\cite{edk2} \href{https://github.com/tianocore/edk2/blob/6c8dd15c4ae42501438a525ec41299f365f223cb/MdePkg/Library/BasePeCoffLib/BasePeCoff.c\#L843-L848}
        {BasePeCoffLib/BasePeCoff.c:843-848}
    } --- all callers to this function that access more than one byte from this pointer without additional
    caution may perform out-of-bounds accesses (\href{https://cwe.mitre.org/data/definitions/125.html}
    {\textit{OOB}}). While there is a practice of calling the function twice, where the second call is passed the range's end
    offset\footnote{
        EDK~II~\cite{edk2} \href{https://github.com/tianocore/edk2/blob/6c8dd15c4ae42501438a525ec41299f365f223cb/MdePkg/Library/BasePeCoffLib/BasePeCoff.c\#L1743-L1746}
        {BasePeCoffLib/BasePeCoff.c:1743-1746}
    }, this is very much unintuitive and error-prone. There are
    actual occurrences of \textit{PeCoffLoaderImageAddress}-based OOB accesses\footnote{
        EDK~II~\cite{edk2} \href{https://github.com/tianocore/edk2/blob/6c8dd15c4ae42501438a525ec41299f365f223cb/MdePkg/Library/BasePeCoffLib/BasePeCoff.c\#L1538-L1542}
        {BasePeCoffLib/BasePeCoff.c:1538-1542}
    }. Furthermore, HII section lookup may access elements without ensuring their prior existence\footnote{
        EDK~II~\cite{edk2} \href{https://github.com/tianocore/edk2/blob/6c8dd15c4ae42501438a525ec41299f365f223cb/MdePkg/Library/BasePeCoffLib/BasePeCoff.c\#L1583}
        {BasePeCoffLib/BasePeCoff.c:1583}

        EDK~II~\cite{edk2} \href{https://github.com/tianocore/edk2/blob/6c8dd15c4ae42501438a525ec41299f365f223cb/MdePkg/Library/BasePeCoffLib/BasePeCoff.c\#L1600}
        {BasePeCoffLib/BasePeCoff.c:1600}
    }. It is also not explicitly documented that requests for data in the TE header are not supported due to the \textit{TeStrippedOffset} subtraction.\\
    This defect is addressed by the raw file and loaded image models~(\ref{sect:imagemodels}).
    
    \item \textbf{Integer wraparounds}
    
    The EDK~II Image Loader is affected by multiple types of \href{https://cwe.mitre.org/data/definitions/190.html}
    {integer wraparounds}. Some of them are likely harmless\footnote{
        EDK~II~\cite{edk2} \href{https://github.com/tianocore/edk2/blob/6c8dd15c4ae42501438a525ec41299f365f223cb/MdePkg/Library/BasePeCoffLib/BasePeCoff.c\#L139}
        {BasePeCoffLib/BasePeCoff.c:139}
    } because they are implicitly accounted for shortly after\footnote{
        EDK~II~\cite{edk2} \href{https://github.com/tianocore/edk2/blob/6c8dd15c4ae42501438a525ec41299f365f223cb/MdePkg/Library/BasePeCoffLib/BasePeCoff.c\#L162-L174}
        {BasePeCoffLib/BasePeCoff.c:162-174}
    }, yet this cannot be allowed as it makes manual review much harder and is error-prone regarding future
    refactoring. There also are some that may cause an infinite loop upon facing unlucky
    values\footnote{
        EDK~II~\cite{edk2} \href{https://github.com/tianocore/edk2/blob/6c8dd15c4ae42501438a525ec41299f365f223cb/MdePkg/Library/BasePeCoffLib/BasePeCoff.c\#L701}
        {BasePeCoffLib/BasePeCoff.c:701}~(for 32-bit $UINTN$)
    }.\\
    This defect is addressed by the raw file and loaded image models~(\ref{sect:imagemodels}).
    
    \item \textbf{Alignment requirement violations}
    \label{funvul:alignment}
    
    Any CPU architecture may impose alignment requirements for data access. Prominent examples include
    x86~(SSE), ARM, MIPS and PowerPC, of which the first two are officially supported by the UEFI right now. While
    most unsupported unaligned accesses result in exceptions, some, e.g. ARMv6 and below\footnote{
        Supported as per UEFI specification~\cite{uefispec}, 2.3.5 AArch32 Platforms
    }, may yield unpredictable behaviour\footnote{
        ARMv6-M\cite{armv6spec}, A3.5.5 Memory access restrictions
    }. The EDK~II
    Image Loader does not verify alignment requirements of offset-based pointers\footnote{
        EDK~II~\cite{edk2} \href{https://github.com/tianocore/edk2/blob/6c8dd15c4ae42501438a525ec41299f365f223cb/MdePkg/Library/BasePeCoffLib/BasePeCoff.c\#L1273}
        {BasePeCoffLib/BasePeCoff.c:1273}
    }. In fact EDK~II does not provide any way to do so at the time of writing by neither allowing usage of
    the \textit{\_Alignof} operator that is part of the C Programming Language~\cite{cstd} nor providing a macro of its own.
    
    Furthermore, there are occurrences of aligned access to data that may legitimately be unaligned\footnote{
        EDK~II~\cite{edk2} \href{https://github.com/tianocore/edk2/blob/6c8dd15c4ae42501438a525ec41299f365f223cb/MdePkg/Library/BasePeCoffLib/BasePeCoff.c\#L1060}
        {BasePeCoffLib/BasePeCoff.c:1060}
    }.\\
    This defect is addressed by hardened design~(\ref{sect:hardeneddesign}) and the raw file and loaded image models~(\ref{sect:imagemodels}).

    \item \textbf{Uninitialized destination bytes}

    The EDK~II Image Loader does not initialize the destination buffer sufficiently\footnote{
        \href{https://bugzilla.tianocore.org/show_bug.cgi?id=1999}{bugzilla.tianocore.org/show\_bug.cgi?id=1999}
    }. This means that the destination area may contain arbitrary bytes that are not covered by the image
    signature. Bugs in the image code, the Image Loader, or similar spots may lead to an unexpected attack vector.
    As the problem areas are part of the image memory, tools will most likely have trouble detecting this.\\
    This defect is addressed by the loaded image model~(\ref{sect:imagemodels}).

    \item \textbf{Function specification violation}

    The Image Loader currently reports success unconditionally when image relocation is requested but
    relocation information has been stripped. Especially it is not verified whether the destination
    address matches the preferred image address\footnote{
        EDK~II~\cite{edk2} \href{https://github.com/tianocore/edk2/blob/6c8dd15c4ae42501438a525ec41299f365f223cb/MdePkg/Library/BasePeCoffLib/BasePeCoff.c\#L923-L931}
        {BasePeCoffLib/BasePeCoff.c:923-931}
    }. This violates the function specification which states the image has been relocated when success is
    returned\footnote{
        EDK~II~\cite{edk2} \href{https://github.com/tianocore/edk2/blob/6c8dd15c4ae42501438a525ec41299f365f223cb/MdePkg/Include/Library/PeCoffLib.h\#L248}
        {PeCoffLib.h:248}
    }.\\
    This defect is addressed by ACSL models.

    \item \textbf{Runtime relocation is optimistic}

    The PE format has no concept to support the relocation processing UEFI needs to perform when
    entering OS Runtime\footnote{
        UEFI specification~\cite{uefispec}, 8.4 Virtual Memory Services
    }. Yet UEFI requires images to be relocated after they have already been executed. This
    may change values at offsets targeted by Base Relocations. The EDK~II Image Loader solves this
    issue by optimistic bookkeeping as to changed values are skipped\footnote{
        EDK~II~\cite{edk2} \href{https://github.com/tianocore/edk2/blob/6c8dd15c4ae42501438a525ec41299f365f223cb/MdePkg/Library/BasePeCoffLib/BasePeCoff.c\#L1815}
        {BasePeCoffLib/BasePeCoff.c:1815}
    }. However, the PE format does not consider loader behaviour of this kind and thus must be
    restricted to not accidentally compromise security.\\
    This defect is addressed by ACSL models.

    \item \textbf{HII section lookup may malfunction}
    
    The current implementation of the HII section lookup may cause unexpected behaviour by unintentionally
    exchanging the loop object (``ResourceDirectory'') during the loop execution\footnote{
        EDK~II~\cite{edk2} \href{https://github.com/tianocore/edk2/blob/6c8dd15c4ae42501438a525ec41299f365f223cb/MdePkg/Library/BasePeCoffLib/BasePeCoff.c\#L1549}
        {BasePeCoffLib/BasePeCoff.c:1549}
    }.\\
    This defect is addressed by ACSL models.

    \item \textbf{Sections might overlap}
    \label{funvul:sectoverlap}
    
    In theory sections may refer to overlapping bytes. This would not be a problem with an
    algorithm hashing the entire binary at once but the Microsoft Authenticode algorithm hashes every section
    individually, which means bytes may be hashed multiple times. This fact can be abused to effectively hash
    a dramatically higher amount of bytes in total than if overlapping sections were prohibited. While no
    evidence of such an attack vector seems to be available at this point, it cannot be taken out of consideration as a
    viable threat in the future.\\
    This defect is addressed by hardened design~(\ref{sect:hardeneddesign}) and ACSL models.

    \item \textbf{TE sections or header may be loaded unaligned}
    \label{funvul:teunaligned}

    For TE images the Image Loader locates virtual addresses by subtracting the TE stripped size first\footnote{
        EDK~II~\cite{edk2} \href{https://github.com/tianocore/edk2/blob/6c8dd15c4ae42501438a525ec41299f365f223cb/MdePkg/Library/BasePeCoffLib/BasePeCoff.c\#L858}
        {BasePeCoffLib/BasePeCoff.c:858}
    }. Because those fields remain untouched by the PE to TE conversion, this results in an address not
    aligned by the image section alignment. This behaviour is not documented by the Image Loader itself,
    however, the \href{https://github.com/tianocore/edk2/tree/6c8dd15c4ae42501438a525ec41299f365f223cb/MdeModulePkg/Core/Pei}
    {\textit{PEI Core}} works around this by adding the TE stripped size to the destination address\footnote{
        EDK~II~\cite{edk2} \href{https://github.com/tianocore/edk2/blob/6c8dd15c4ae42501438a525ec41299f365f223cb/MdeModulePkg/Core/Pei/Image/Image.c\#L384-L395}
        {Core/Pei/Image/Image.c:384-395}
    }. This in return results in the TE image header not being loaded at an address aligned by the image
    section alignment. The \href{https://github.com/tianocore/edk2/tree/6c8dd15c4ae42501438a525ec41299f365f223cb/MdeModulePkg/Core/Dxe}
    {\textit{DXE Core}} does not attempt to work around this at all.\\
    This is especially bad for eXecute In Place (XIP) images as they are not loaded at all and thus remain unaligned in the
    flash memory. However, XIP images are out of the scope of this document.\\
    This defect is addressed by the loaded image model~(\ref{sect:imagemodels}).

    \item \textbf{Non-conformant MS-DOS Stub is tolerated for TE}
    \label{funvul:tedos}

    According to the TE specification \textit{StrippedSize} bytes are stripped from the start of the file
    before the TE header is added\footnote{
        PI specification~\cite{pispec}, Volume 1: PEI Core, 15.2 PE32 Headers, TE Header
    }. This implies there must not be any additional headers preceding the TE header. EDK~II however supports
    a preceding MS-DOS Stub for TE images in the Image Loader\footnote{
        EDK~II~\cite{edk2} \href{https://github.com/tianocore/edk2/blob/6c8dd15c4ae42501438a525ec41299f365f223cb/MdePkg/Library/BasePeCoffLib/BasePeCoff.c\#L79-L129}
        {BasePeCoffLib/BasePeCoff.c:79-129}
    }.\\
    This defect is addressed by the loaded image model~(\ref{sect:imagemodels}).

    \item \textbf{TE \textit{SizeOfHeaders} is inadequate}
    \label{funvul:tesize}
    
    EDK~II defines \textit{SizeOfHeaders} for TE images via \textit{BaseOfCode}\footnote{
        EDK~II~\cite{edk2}  \href{https://github.com/tianocore/edk2/blob/6c8dd15c4ae42501438a525ec41299f365f223cb/MdePkg/Library/BasePeCoffLib/BasePeCoff.c\#L139}{BasePeCoffLib/BasePeCoff.c:139}
    } for no obvious reason.
    Unfortunately \textit{SizeOfHeaders} cannot be reconstructed precisely from the TE image, however
    \textit{BaseOfCode} may cause non-obvious issues with images that have an unexpected section order.\\
    This defect is addressed by ACSL models. It additionally needs to be addressed on PI specification~\cite{pispec} level.
\end{enumerate}

\subsection{Unaddressed considerations}
\begin{enumerate}
    \item \textbf{Base Relocations might overlap}
    
    Theoretically Base Relocations may refer to overlapping bytes. While this obviously is unreasonable,
    the PE format~\cite{pespec} does not cover such a possibility or how to avoid it. As there is no guarantee
    of their order or similar, expensive bookkeeping would be required to ensure all Base Relocations refer to
    disjoint ranges. Accounting for this does not seem to be beneficial as all the operations are memory-safe and overall safety cannot be guaranteed in either case.

    \item \textbf{Overcomplicated hashing algorithm}
    \label{consid:ComplicatedHash}
    
    The PE hashing algorithm specified by the Authenticode format\footnote{
        Authenticode~\cite{pehashspec}, Calculating the PE Image Hash
    } has been designed for a debugging-friendly in-OS usage, mostly to allow file modifications after the
    signing process. However, especially in security-critical low-level software, this is
    not permitted. Instead, it would be more advisable to use a simpler hashing algorithm that hashes the file
    entirely~(manifest-based verification) or till the trailing signature. The PE format guarantees the
    certificate and signature information are indeed trailing\footnote{
        PE format~\cite{pespec},
        \href{https://docs.microsoft.com/en-us/windows/win32/debug/pe-format\#the-attribute-certificate-table-image-only}
        {The Attribute Certificate Table~(Image Only)}
    }, so the most intuitive way to hash the file is to hash all bytes from the beginning of the file to the
    start of this data --- the related security directory information must be set before the image is hashed.
    
    \item \textbf{Hashing as early as possible}
    
    Please refer to ``Overcomplicated hashing algorithm''~(\ref{consid:ComplicatedHash})
    for context. It is common practice to hash a binary and validate its signature as early as possible in
    the loading process to avoid abusing loader bugs. However, since PE has a very complex hashing
    algorithm many of the image properties need to be verified beforehand. Therefore, it is unreasonable to
    postpone the rest of the validation.
    
    \item \textbf{Image fields not covered by the hash}
    
    The Authenticode algorithm for PE image hashing dictates skipping the checksum and the security directory
    information of the image header\footnote{
        Authenticode~\cite{pehashspec}, Calculating the PE Image Hash
    }. Good security practice suggests that these fields shall be zero in the destination area. Meanwhile,
    alternative algorithms have surfaced, such as in Mac EFI where hashing operates on the entire file till
    the signature information. In this case those fields must be loaded into the destination area.
    Considering the usage of an alternative hashing algorithm is strongly recommended, zeroing the three
    affected fields is likely not worth it. Furthermore, the new design allows omitting loading the headers
    explicitly and at the same time allows them to be covered by a section, both of which do not require such
    actions.
\end{enumerate}

\section{AstraVer Toolset}
\label{sec:astraver}

Formal verification is one of the software verification techniques commonly utilised in safety-critical
software development lifecycles in aerospace, automotive, medical, energetic, and other
similarly high-risk industries.
The aim of formal verification is to mathematically prove the correspondence of an algorithm to its
formal specification in order to be able to reason about the safety and correctness of the implementation
and reducing the impact of a human error.

Although formal verification is generally considered heavyweight and time-consuming, modern tools involving
various verification strategies with automated reasoning make it possible to verify real-world software.
Among the well-known success stories are avionics software verification in Airbus~\cite{airbus} done
primarily with Caveat and Frama-C, Astra Linux security module verification with the AstraVer plugin for
Frama-C~\cite{avpaper}, Hyper-V hypervisor verification in Microsoft with VCC~\cite{hyperv}, seL4 microkernel
in OK Labs with Isabelle/HOL~\cite{sel4}, and CompCert C compiler at INRIA with Coq\cite{compcert}.

In general, there are two formal verification approaches:

\begin{itemize}
\item Synthesising program code from the formal proof.
\item Verifying already written program code.
\end{itemize}

The choice between the two strongly depends on the nature of the target software and the ability to modify it.
Synthesising a formally verified program commonly allows proving larger software or software with complex
mathematical algorithms such as High Assurance Cryptographic Library~(HACL*)~\cite{hacl-star}. On the other
side verifying an existing program allows for uninterrupted development by separating the deadlines
of a working prototype and fully verified software in the lifecycle and enables the ability to integrate
legacy and third-party software without the need to rewrite it from the ground up. When verifying large codebases,
the general approach is to only write formal proofs for most critical areas and use other means of software verification for the other making the process more affordable.

Although no test suite can provide the guarantees comparable to formal verification simply because the number
of tests is always finite, there still are various methods that have proven to be effective in safety and
security areas, like tests with full MC/DC coverage or fuzzing. These methods should not be neglected
and may be used together with formal verification. It should always be taken into account that while we
significantly improve software reliability by providing a formal proof, mistakes can still happen within
the proof, the proving tool itself, or on other layers like in the compiler or even in the hardware.

Since the aim of this project was to demonstrate the ability to create formal proofs for existing
software in the UEFI environment, we chose the AstraVer Toolset. This tool allows expressing function
behaviour as a specification written in a dedicated language called ACSL~\cite{acsl}
consisting of a precondition and a postcondition. The tool translates annotated C code into a set of logical
formulae (verification conditions), the general validity of which is equivalent to program correctness as in
Floyd-Hoare logic.

The choice of the AstraVer Toolset over the base Frama-C version was dictated by the wider set of
abilities required for proving real-world software with pointer operations, bitwise arithmetic,
type inspection, and the like. In this sense AstraVer Toolset is a further development of the Jessie
plugin~\cite{jessie} for Frama-C tested on existing system software such as the Linux kernel.
It implements a new memory model~\cite{avmemmodel} that allows to support the \textit{container\_of} construct,
pointer type reinterpretation between integer types, including types of different size, bitwise
arithmetic operations on expression-level, and has several other features~\cite{avpaper}.

\section{Implementing the Image Loader}
\label{sec:implementation}

\subsection{Memory model}

We assume the CompCert Memory Model v2~\cite{compmem2} as a base for the proof design. Punctually its guarantees are relaxed towards the previous CompCert Memory Model v1 as it is closer to the model AstraVer uses internally\footnote{
    Linux kernel verification\cite{avpaper}, 4 Region separation in Jessie,
    
    5.1 Jessie byte-level block memory model
}. Namely:
\begin{itemize}
  \item For functions performing only read operations valid pointers of different types are not required to refer to disjoint memory regions from valid byte arrays.
  \item For functions performing direct write operations valid pointers of different types must all refer to disjoint memory. Most notably no memory may overlap between a byte array and a different data type. This must either propagate to the caller, or the \textit{assigns} clause must include all representations (e.g. the full byte range covered by a modified data structure), or there must be some obvious internal proof the data did not change.
\end{itemize}

The reason for this is that to AstraVer valid pointers of different types always point to disjoint memory. For reading operations this is acceptable. If a write action is performed to a valid pointer however, the change will not be reflected for any other valid pointer of a different type partially or fully aliasing the same memory rendering the proof inaccurate.

For the production environment, we fully assume the v2 model to introduce performance improvements such as reading aligned 32-bit values in one operation. To accomplish this, we use macros that in the proof environment operate on byte-level but still have correct alignment constraints.

\subsection{Type model}

\subsubsection{Data type alignment model}
\label{sect:alignmodel}

According to the UEFI specification, all fundamental types are aligned naturally. For structures it declares they are aligned by the maximum size out of all internal data\footnote{
    UEFI~\cite{uefispec}, 2.3.1 Data Types
}, but real-world compilers use the maximum internal alignment instead. Natural alignment in this context refers to alignment being equal to the minimum of the data type size and the maximum alignment size which is defined by the architecture's UEFI ABI. Please remember that floating-point types are not supported by UEFI.

These assumptions were used to formulate an acceptable alignment model in the lack of a tool-supported one:
\begin{align*}
    A\_MAX  &:=
    \left\{
        \begin{array}{ll}
            4  & \mbox{for IA32}\\
            8  & \mbox{otherwise}\\
        \end{array}
    \right. \\
    \_Alignof(T) &\defn \min\{ sizeof(T), A\_MAX \}
\end{align*}

For fundamental types and the UEFI ABI this definition is accurate. For aggregate types and unions however, the alignment our type model yields is often too high as the type size is often larger than the maximum alignment of each internal datum. Such alignment proofs are considered ``proof of concept'' rather than an actual part of the formal verification, and we must admit our current toolset does not allow for an accurate result. A new revision of the AstraVer Toolset is currently in development and the mentioned culprits are considered for its design.

We addressed this limitation as such for the moment. When an operation or an ACSL annotation lead to a pointer known to be aligned for one type also be aligned for the other, a $\_Static\_assert$ is used to prove transitivity in case not all distinct types are explicitly verified.

Considering the data type definitions and their sizes, data type alignments can only ever be a power of two. Powers of two especially satisfy the following condition of the C Programming Language standard, which we used in multiple spots for code optimisations:
``When an alignment is larger than another it represents a stricter alignment.''\footnote{
    C17~\cite{cstd}, 6.2.8.7 (unchanged since C11)
}\\

\subsubsection{Pointer target alignment model}

The pointer target alignment model is incomplete but sufficient for all required use cases. Any memory allocation function declares a guarantee for the alignment of the returned pointer~(e.g. 4 KB alignment for page-wise allocation or 8-byte alignment ($A\_MAX$) for $AllocatePool$). From there, additions to these pointers are axiomatically modelled to yield a pointer of alignment $a$ if the pointer ($p$) and offset ($o$) are both aligned by $a$.
\begin{gather*}
    \forall p,a,o: aligned(p,a) \land o \bmod a = 0 \Rightarrow aligned(p+o,a)
\end{gather*}

This is correct as per the data type alignment model as the sum of two values divisible by a power of two itself is also divisible by that power of two.\\

\subsubsection{Pointer target validity model}

The C Programming Language does not have a standard-compliant way to identify the type of data structures from the in-memory representation. While we cannot give a definition for the sufficient conditions to determine validity, we can define it based on the required ones of both the C Programming Language standard and the PE format:

For all the file and image locations the PE format designates a data structure for, it is valid if and only if the entirety of its size is contained in the bounds of the file or image and the location satisfies the alignment requirements of the designated data type.\\

\subsubsection{Fundamental type definitions}
\label{subsec:funtypes}

In the following we define the utilised fundamental data types of our Image Loader derived from the UEFI specification\footnote{
    UEFI~\cite{uefispec}, 2.3.1 Data Types
}.\\

From the generic definitions
\begin{align*}
    FALSE   & := 0, TRUE := 1\\
    UINT(x) & := \{ n \in \mathbb{N} \mid 0 \leq n < 2^x \}
\end{align*}
we define the generic fundamental types. Please note that for simplicity, we assume that $CHAR8$ is unsigned.
\begin{center}
    \begin{tabular}{ |c|c|c|c| } 
        \hline
        Type      & Definition       & sizeof & \_Alignof\\
        \hline
        $BOOLEAN$ & $\{TRUE,FALSE\}$ & 1      & 1\\
        $CHAR8$   & $UINT(8)$        & 1      & 1\\
        $UINT8$   & $UINT(8)$        & 1      & 1\\
        $UINT16$  & $UINT(16)$       & 2      & 2\\
        $UINT32$  & $UINT(32)$       & 4      & 4\\
        $UINT64$  & $UINT(64)$       & 8      & A\_MAX\\
        \hline
    \end{tabular}
\end{center}
\begin{gather*}
    UINTN :=
    \left\{
        \begin{array}{ll}
            UINT32 & \mbox{for IA32, ARM}\\
            UINT64 & \mbox{for X64, AArch64, RISC-V}\\
        \end{array}
    \right.
\end{gather*}
\\
\subsubsection{Aggregate types and unions}

The composition of fundamental data types and recursively aggregate types and unions follows the C Programming Language. Some constructs cannot be expressed in the C Programming Language, such as arrays of variably-sized data --- other means of modelling such as regular expressions will be used in the following to express their composition. Data structures modelled in the referenced specifications, such as the UEFI or the PE specifications, that are expressible in the C Programming Language are assumed to be known and will not be explicitly modelled.\\

\subsection{Proof definition}
\label{sect:imagemodels}

\subsubsection{Model-aided proof goals}

To increase the security and stability of the Image Loader, we specified the following as our verification requirements. In particular, they include considerations in context of the C Programming Language.\\
\textit{Safety requirements}:
\begin{itemize}
    \item Non-modulo integer arithmetic does not wrap around.
    \item All functions terminate on all external inputs.
    \item Memory accesses happen in valid bounds and are aligned.
\end{itemize}
\textit{Functional requirements}:
\begin{itemize}
    \item Raw files that do not conform to the modelled subset of the PE or TE file formats are discarded.
    \item The image is loaded correctly and the result is deterministic (please refer to design defect \ref{desvul:nondet}).
    \item Data expected to remain valid is not invalidated (e.g. applying Base Relocations does not modify the Relocation Directory itself, please refer to design defect \ref{desvul:reloctrashing}).
    \item Individual Base Relocations are applied correctly.
\end{itemize}

The safety of integer operations and function termination are automatically proved by AstraVer unless specified otherwise. No such conditions have been defined for our Image Loader, hence both goals are always met.

To satisfy the remaining requirements, we will define predicates based on the PE and TE specifications, as well as our own constraints that compose the file and image models (``format model''). They are translated to ACSL definitions and the code is annotated in such a way that non-conformant inputs abort the process.\\

\subsubsection{Notation}

Below you will find a table of defined operators, functions, acronyms and abstractions.
\begin{center}
    \begin{supertabular}{ |c|c| }
        \hline
        $\#$                    & Amount of elements in a collection\\
        \hline
        \textit{alignBRT}($r$)  & Base Relocation target alignment\\
        \textit{align}($v$,$a$) & The least multiple of $a$ not less than $v$\\
        \textit{size}($d$)      & Size of the variably-sized data structure\\
        \textit{sizeBRT}($r$)   & Base Relocation target size\\
        \textit{vaBRT}($r$)     & Base Relocation target VA\\
        \textit{typeBRT}($r$)   & Base Relocation target data type\\
        \hline
        $BR$                    & Base Relocation (inner-page)\\
        $BRB$                   & Base Relocation Block (page-wise)\\
        $BS$                    & Block size\\
        $DOS$                   & MS-DOS Stub\\
        $H$                     & Raw file headers\\
        $O$                     & Offset\\
        $PE32$                  & COFF and PE32 Optional Header\\
        $PE32Plus$              & COFF and PE32+ Optional Header\\
        $RD$                    & Relocation directory\\
        $RS$                    & Size of raw file data\\
        $S$                     & Section Table ($S = (SH)^+$)\footnotemark\\
        $SH$                    & Section header\\
        $T$                     & Type\\
        $TE$                    & TE Header\\
        $VA$                    & Virtual address\\
        $VS$                    & Virtual size\\
        \hline
        $c.FS$                  & Raw file size\\
        $c.HS$                  & Raw file \textit{SizeOfHeaders}\\
        $c.IS$                  & Image \textit{SizeOfImage}\\
        $c.RDS$                 & Image Relocation Directory size\\
        $c.RDV$                 & Image Relocation Directory VA\\
        $c.SA$                  & Image \textit{SectionAlignment}\\
        $c.SO$                  & Raw file Section Table offset\\
        \hline
    \end{supertabular}
\end{center}
\footnotetext{
    The requirement of at least one section is imposed by us.
}

The operators not mentioned in the table are defined as in the C Programming Language standard.
Please remember that $sizeof$ does not include the size of flexible arrays. 
$c$ is used in the following to denote an image context, i.e. image loader state structure,
to reason in an image type agnostic fashion.\\
\hfill

\subsubsection{File headers}
\label{subsubsec:hdrmodel}

As previously described, the memory model allows for overlapping regions of different data types for read operations. The raw file is read-only throughout the library, thus no considerations regarding overlapping need to be made. We require all file buffers to be aligned by the maximum fundamental alignment ($A\_MAX$). Please note that TE stripping will not be considered for simplicity.
\begin{align*}
    PE & = PE32|PE32Plus\\
    H  & = (TE|[DOS] \circ\textsuperscript{\ref{foot:circ}} PE)\textsuperscript{\ref{foot:pad}}
\end{align*}
\stepcounter{footnote}\stepcounter{Hfootnote}
\footnotetext{
    \label{foot:circ}
    The $\circ$ operator denotes concatenation.
}
\stepcounter{footnote}\stepcounter{Hfootnote}
\footnotetext{
    \label{foot:pad}
    PE32, PE32Plus, and DOS data structures may be succeeded by arbitrary

    data or padding which is considered to be part of its definition for simplicity.
    
    This trailing data is sized to hold the alignment for the following data.
}
There are several constraints, especially for size fields. However, they will not be modelled for being obvious from the specifications. Due to the flexible format of the headers, formal definitions will be omitted, but the following must hold:
\begin{itemize}
    \item The PE header offset from the MS-DOS Stub must be aligned.
    \item $size(H) \leq c.FS$
\end{itemize}

The initialization routine verifies the input header and succeeds if and only if it matches the definition of $H$. This allows us to prove the following:
\begin{itemize}
    \item The file headers are all in bounds and aligned.
    \item The file headers conform to the format model.
\end{itemize}
\hfill
\subsubsection{File Section Table}

The image's virtual address space is composed of one or more sections which must be sorted in ascending order and be contiguous in terms of section alignment\footnote{
    PE Format~\cite{pespec} ``Section Table (Section Headers)''
}. It is obvious that the loaded sections are disjunct. Their headers reside adjacent to $H$ (thus we have $H \circ S$). For PE32 and PE32+ images they must also fit their \textit{SizeOfImage} value.\\

While the PE format does not explicitly describe the constraints of the first virtual address, most tools set it to the aligned end address of the image headers because they expect the Image Loader to manually load it into the execution environment. However, the headers should not be accessed except by the Image Loader and debugging instruments in UEFI. Thus, we decided to make explicitly loading them optional, to prohibit section file data to overlap with them and to allow the first section to be the beginning of the address space. Both allowed values for the first Virtual Address are aligned by \textit{SectionAlignment} and hence all sections are correctly aligned\footnote{
    Malformed binaries reached production due to the current Image Loader
    
    failing to verify those properties. In response, we introduced an optional
    
    mode which does not verify strict continuity or alignment but only the
    
    derived properties such as ascending order and separation of section
    
    memory. Its model is out of scope.
}.
\begin{align*}
    c&orrectSA(c,s) \defn\\
    & \#s > 0 \land (s[0].VA = 0 \lor s[0].VA = align(c.HS, c.SA)) \land\\
    & \bigwedge_{i=1}^{\#s-1} s[i].VA = align(s[i-1].VA + s[i-1].VS, c.SA) \land\\
    & align(s[\#s-1].VA + s[\#s-1].VS, c.SA) \leq c.IS
\end{align*}

Every section has an offset into the file buffer at which its data is located as well as its size --- obviously, the file bounds must be respected. Please note that the size that will be copied by the loading code for every section header $sh$ is $\min{sh.VS, sh.RS}$ and $[sh.VA + sh.RS, sh.VA + sh.VS)$ is filled with zeros.
\begin{align*}
    v&alidMemS(c,s) \defn\\
    & c.SO \bmod \_Alignof(s[0]) = 0\textsuperscript{\ref{foot:shaligned}} \quad\land\\
    & c.SO + \#s \cdot sizeof(s[0]) \leq c.FS \quad\land\\
    & \bigwedge_{sh \in s} 0 < sh.RS \Rightarrow c.HS \leq sh.O \land sh.O + sh.RS \leq c.FS
\end{align*}
\stepcounter{footnote}\stepcounter{Hfootnote}
\footnotetext{
    \label{foot:shaligned}
    This constraint is our own and follows from the type model.
}
The constraints regarding file and image memory add up to the section correctness.
\begin{gather*}
    correctS(c,st) \defn correctSA(c,st) \land validMemS(c,st)
\end{gather*}

The initialization routine verifies the input Section Table and succeeds if and only if they satisfy $correctS$. This allows us to prove the following:
\begin{itemize}
    \item The section headers are all in bounds and aligned.
    \item The section headers conform to the format model.
    \item The raw and virtual section targets are all in bounds (thus loading does not cause OOB).
    \item The virtual address space is contiguous and thus deterministic.
    \item Loading from the file buffer to the image memory is injective (thus loading a section does not invalidate data loaded by previous sections).
\end{itemize}
\hfill
\subsubsection{Image Relocation Directory}
\label{subsubsec:relocmodel}

The Relocation Directory is a concatenation of arbitrarily many Base Relocation Blocks.
\begin{gather*}
    RD = (BRB)^*
\end{gather*}

We define $size()$ functions to be able to clearly express their bounds.
\begin{align*}
    size(brb) & \defn sizeof(BRB) + \#brb.BR \cdot sizeof(brb.BR[0])\\
    size(rd) & \defn \sum_{brb \in rd.BRB}size(brb)
\end{align*}

Base Relocation targets vary in target data type, data size and data alignment requirements per type. For the scope of this project, we only allow a subset of the Base Relocation types found in the real world. They are characterized as such (where $C8$ and $U32$ denote $CHAR8$ and $UINT32$):
\begin{align*}
    typeBRT(br) & \defn
    \left\{
        \begin{array}{ll}
            C8[4]  & \mbox{if } br.T = HIGHLOW\\
            C8[8]  & \mbox{if } br.T = DIR64\\
            U32[2] & \mbox{if } br.T = ARM\_MOV32T\\
            VOID   & otherwise
        \end{array}
    \right.\\
    alignBRT(br) & \defn
    \left\{
        \begin{array}{ll}
            1      & \mbox{if } br.T = HIGHLOW\\
            1      & \mbox{if } br.T = DIR64\\
            4      & \mbox{if } br.T = ARM\_MOV32T\textsuperscript{\ref{foot:mov32talign}}\\
            \infty & otherwise
        \end{array}
    \right.\\
    sizeBRT(br) & \defn
    \left\{
        \begin{array}{ll}
            4      & \mbox{if } br.T = HIGHLOW\\
            8      & \mbox{if } br.T = DIR64\\
            8      & \mbox{if } br.T = ARM\_MOV32T\\
            \infty & otherwise
        \end{array}
    \right.\\
    vaBRT(brb,br) & \defn brb.VA + br.O
\end{align*}
\stepcounter{footnote}\stepcounter{Hfootnote}
\footnotetext{
    \label{foot:mov32talign}
    This requirement is not specified by the PE format, but is derived from
    
    the ARM architectural requirements.
}
For the target of a Base Relocation, its start must be correctly aligned for its type and its range must not overlap with the Relocation Directory memory. This particularly satisfies the memory model condition of aliased data (the byte and the data structure representations of the Relocation Directory) being constant.
\begin{alignat*}{2}
    t& := vaBRT(brb,br), s := sizeBRT(br)\\
    c&orrectRT(c,brb,br) \defn \\
    & t + s \leq c.IS \land t \bmod alignBRT(br) && = 0 \quad \land\\
    & [t, t + s) \cap [c.RDV, c.RDV + c.RDS) && = \emptyset
\end{alignat*}
\\
With this, we can define the remaining correctness.
\begin{align*}
    c&orrectBR(c,brb,br) \defn\\
    & sizeBRT(r) < \infty \land correctRT(c,brb,br)\\
    c&orrectBRB(c,brb) \defn\\
    & brb.BS \bmod \_Alignof(BRB)\textsuperscript{\ref{foot:brbalign}} = 0\textsuperscript{\ref{foot:brbpad}} \quad\land\\
    & brb.BS = size(brb) \land  \forall r \in brb.BR: correctBR(c,brb,br)\\
    c&orrectRD(c,rd) \defn\\
    & c.RDV \bmod \_Alignof(BRB) = 0 \quad\land\\
    & c.RDV + c.RDS \leq c.IS \land c.RDS = size(rd) \quad\land\\
    & \forall brb \in rd.BRB: correctBRB(c,brb)
\end{align*}
\stepcounter{footnote}\stepcounter{Hfootnote}
\footnotetext{
    \label{foot:brbalign}
    The PE format explicitly requires 32-bit alignment which however is
    
    equivalent to $\_Alignof(BRB)$ on all platforms.
}
\stepcounter{footnote}\stepcounter{Hfootnote}
\footnotetext{
    \label{foot:brbpad}
    This imposes a constraint on $\#brb.BR$. The $ABSOLUTE$ Base
    
    Relocation type is used to pad Base Relocation Blocks to an aligned size.
}
The relocation routine verifies the input Relocation Directory and succeeds if and only if it satisfies $correctRD$. We require Base Relocations to be processed in the order of their appearance. This allows us to prove the following:
\begin{itemize}
    \item The Relocation Directory and all its targets are in bounds and aligned.
    \item The Relocation Directory conforms to the format model.
    \item Relocation does not invalidate data used by the loader (most notably, applying Base Relocations does not modify the Relocation Directory itself).
    \item The resulting memory is deterministic for each distinct load address.
\end{itemize}

A proof of the correct application of individual $HIGHLOW$ and $DIR64$ relocations is available. However, due to the lack of a guarantee that the targets do not overlap, modelling the result would have been too expensive. Application effects beyond invalidation are out of scope for this document.\\
\hfill

\subsubsection{Model-aided proof results}

The format model allows us to prove all previously defined goals as can be seen in sections \ref{subsubsec:hdrmodel} to \ref{subsubsec:relocmodel}. However, please note the following:
\begin{itemize}
    \item The image model only mostly covers the image data structures and does not cover algorithms at all. Please refer to the Image Loader codebase.
    \item Only data that is used by the Image Loader is modelled and validated. Any uninvolved data may be invalid.
    \item The defined alignment model is not sufficient to conform to the C Programming Language (for aggregate types and unions). Additional means of ensuring correctness were taken such as manual code review, introduction of $\_Static\_assert$ usages and dynamic testing.
    \item While all core code was proven for safety, optional code such as ARM or RISC-V Base Relocations, the Runtime relocation bookkeeping buffer and the hashing algorithm were not proven for correct functionality. Instead, all the optional code was manually reviewed for correctness due to the proving efforts required.
    \item Code beyond the scope of the core Image Loader such as explicit debug directory loading and HII resource section lookup have not been proven at all but went through manual code review and dynamic testing. They are entirely optional and we expect them to be disabled in environments with special security requirements.
\end{itemize}

\subsection{Hardened design}
\label{sect:hardeneddesign}

Beyond proving correctness, several changes have been made to the code design to allow for a more stable and secure operation that satisfies security-critical demands.
\begin{itemize}
    \item Only main system memory is allowed as a data source for raw files~(addresses design defect \ref{desvul:TOCTOU}).
    \item Authenticode hashing is integrated into the core library code~(addresses design defect \ref{desvul:decentralised}).
    \item The hash function works similarly to \textit{HashUpdate}~(addresses design defect \ref{desvul:ciphers}).
    \item Loading the header and debug information is optional~(partially addresses design defect \ref{desvul:infoleak}).
    \item A section discarding API is provided~(partially addresses design defect \ref{desvul:infoleak}).
    \item Uncommon ambiguous Base Relocation types are not supported~(addresses design defect \ref{desvul:unclearreloc}).
    \item All public functions that deal with variably-sized structures take size as a parameter or output its maximum or exact size~(addresses design defect \ref{desvul:OOB}).
    \item Public context reads or writes are not part of the library design~(addresses design defect \ref{desvul:funcontr} and partially addresses design defect \ref{desvul:ambiguousctx}).
    \item Public functions are made available for all tasks a caller needs to perform~(partially addresses design defect \ref{desvul:ambiguousctx}).
    \item All public functions return a result that can indicate errors~(addresses design defect \ref{desvul:statusless}).
    \item Authenticode can optionally refuse to hash overlapping raw section data~(addresses functional defect \ref{funvul:sectoverlap}).
    \item ARM and RISC-V Base Relocations are optional independently of the target architecture allowing for more platform flexibility.
\end{itemize}

\section{Conclusion}

Creating secure software is a problem continually addressed by improving software development
life-cycle processes and involving various software verification techniques. Upon
the exploration of the EDK~II codebase, which is a de-facto standard implementation base
for the majority of the modern UEFI firmwares, we were able to identify numerous defects
accompanied by an undesirable track of security issues found in the past. With the entire
UEFI modular architecture being at a risk due to a potential compromise of the EDK~II image
loader we created our own implementation that can be used as a drop-in replacement upstream
with only few changes required.

In order to avoid the issues of the original implementation, we thoroughly analysed the code
and provided a detailed report of the found defects. This review showed that
while some issues were essentially programmer errors that can be resolved by submitting
patches, such as missing bounds checking or memory initialization, several others are design defects, like context ownership violations or TOC/TOU issues with untrusted
storage or networking. The state-of-the-art makes creating a new Image Loader with a similar
interface more practical for verification and maintenance than trying to update the existing one.

During the development of the new UEFI Image Loader we applied industry-standard practices
for security-critical software in both the design and the toolset. We separated the optional
code generally unintended for use in production environment, reworked the ownership semantics,
enabled the use of bounds checking arithmetic, and addressed all the other defects identified
during the analysis stage. Some issues found in the implementation caused by the ambiguous parts
of the UEFI and PE specifications were also addressed based on real-world compiler implementations
and existing images. Both core and optional code underwent static analysis with SVACE
and continual fuzzing with libFuzzer reaching full line and branch coverage, except for the
defensive code.

In order to provide higher reliability guarantees for the core code we created a formal proof
with the help of industry-standard software used for formal verification in safety-critical
systems: Frama-C with the AstraVer plugin. With accordance to the used verification model
the image is discarded when it does not comply to the PE or TE specification, and is correctly
loaded when it does. We proved that the loaded image can also be correctly relocated.
In addition, we proved alignment, arithmetic, memory safety, and function termination.
For the proof we used CVC3, CVC4, and Z3 solvers. To reduce the possibility of a true negative
bug in the solver or its integration, as it happened for us several times with Alt-Ergo,
we ensured that most of the verification conditions are proved by at least two solvers.

Since the UEFI environment does not have hard real-time requirements and does not specify
stack capabilities, the formal proof for worst-case execution time (WCET) and
stack usage are outside the scope of this paper. However, the recursion-free design 
with no variable-length arrays lets us be confident of a possibility of this happening in the future.
For the time being the UEFI watchdog services are an acceptable countermeasure for this class of software.

We tested the implementation on a number of existing UEFI applications and DXE drivers
including Apple macOS, GNU Linux, and Microsoft Windows bootloaders and supplemental
drivers in the firmwares from 2013 to date with no issues discovered. Considering the results, we hope that our
experience encourages the UEFI industry to incorporate this piece of software in their
firmwares and use our approach to provide formal proofs for other critical firmware
components.

\section*{Acknowledgements}

We would like to thank the members of the ISP RAS Linux Verification Center, Mikhail Mandrykin and Alexey Khoroshilov in particular, for invaluable help and feedback on the AstraVer Toolset and general approach. We are also grateful to all our colleagues and the Ivannikov ISP RAS Open committee for their reviews, especially Laszlo Ersek from Red~Hat for incredible patience and a most thorough analysis.


\end{document}